\documentclass[sigconf, nonacm]{acmart}
\AtBeginDocument{%
  }

\acmYear{2026}
\acmConference[RecSys '26]{20th ACM Conference on Recommender Systems, Workshops}{September 28-October 2, 2026}{Minneapolis, USA}

\usepackage{multirow}
\usepackage{algorithm}
\usepackage[noend]{algpseudocode}
\usepackage{amsmath}

\begin{document}

\title{Stageboost: Recommending Signals based on Counterfactual Estimations}

\author{
Darpan Singhal$^{1}$,
Matan Mandelbrod$^{2}$,
Tal Franji$^{2}$,
Manasa Kolla$^{1}$,
Vipul Gaba$^{1}$,
Yuri Brovman$^{3}$
}

\email{{dsinghal, mmandelbrod, tfranji, manak, vgaba, ybrovman}@ebay.com}

\affiliation{
  \institution{$^{1}$eBay}\country{$^{1}$India}
  \institution{$^{2}$eBay}\country{$^{2}$Israel}
  \institution{$^{3}$eBay}\country{$^{3}$USA}
}

\begin{abstract}
Signals are short textual or visual snippets displayed on the eBay View-Item (VI) page, providing additional, contextual information for users about the viewed item. The aim of displaying these signals is to facilitate intelligent purchase and to incentivize engagement. In this paper, we present a 2 stage xgboost based model that optimally populates the VI page with signals. This approach has shown a 0.08\% lift in overall GMB (Gross Merchandise Bought) and 0.58\% increase in Parts and Accessories GMB, primarily due to increase in conversion of high average price items in online experimentation.
\end{abstract}

\maketitle

\section{Introduction}
The View Item (VI) page is one of the most critical touch points in eBay's customer journey, receiving hundreds of millions of daily views. This page serves as the primary information hub for buyers, displaying comprehensive item details including images, price, descriptions, seller information, and shipping options. In addition to these core elements, the VI page features brief, contextual text cues known as signals that are strategically positioned across predefined placements to guide purchase decisions and encourage user engagement.       

\begin{figure}[t]
    \centering
    \includegraphics[
        trim={0cm 0.1cm 0cm 0.1cm},
        clip,
        width=\linewidth
    ]{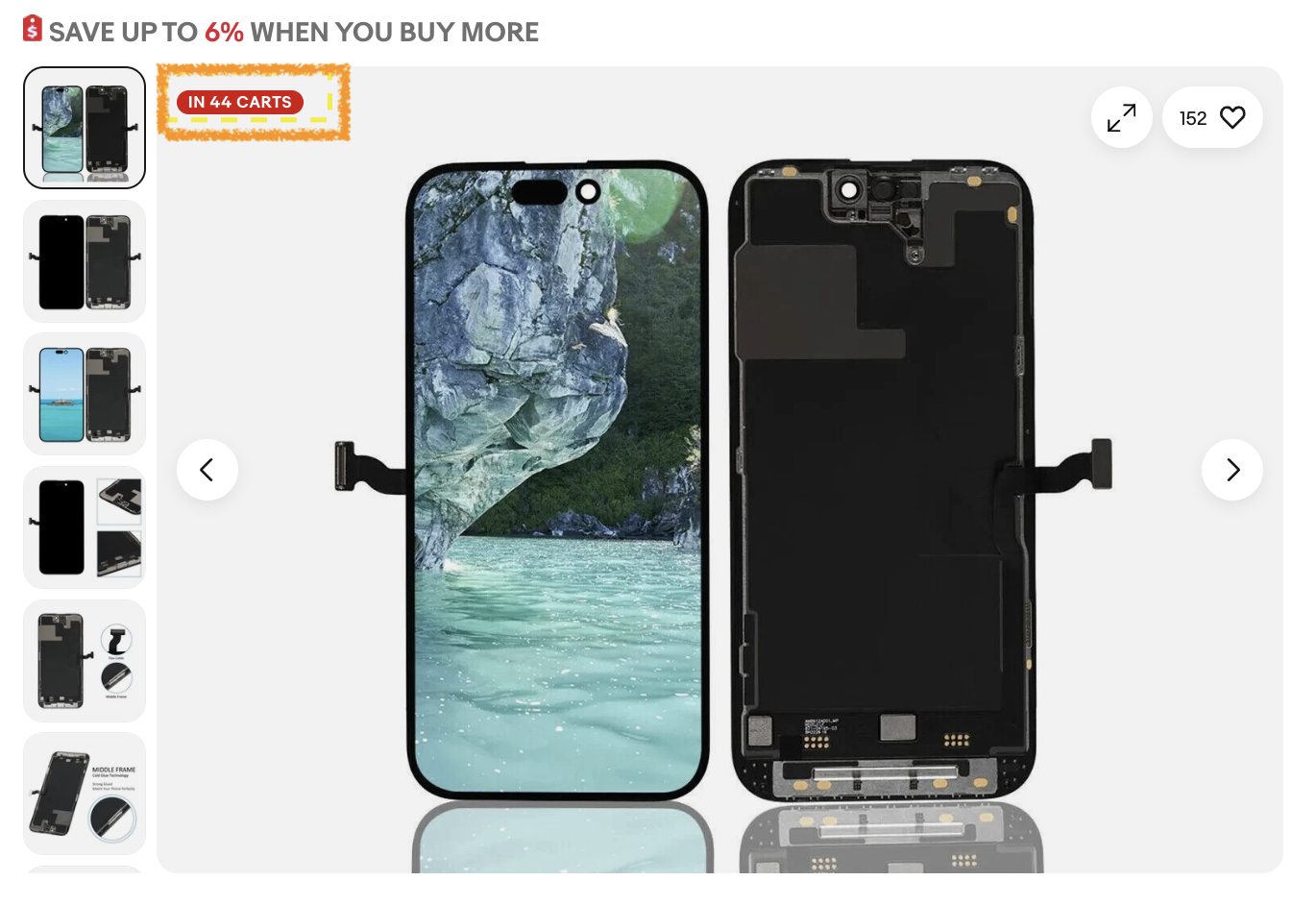}
    \caption{Example of message placement styles, urgency placement in particular.}
    \label{fig:intro}
\end{figure}

Each placement is thematically organized and displays a curated set of candidate signals aligned with its purpose. For instance, the Urgency placement which is prominently positioned over the image panel on mobile devices shows signals that highlight time-sensitive or scarcity-related details, such as “Last one available” or “5 sold in the last 24 hours”. Similarly, the Conversational placement features signals related to item benefits and trustworthiness, such as "Free shipping" or "30-day returns accepted". \autoref{fig:intro} illustrates the placement scheme on the VI mobile experience with examples of displayed signals (marked with orange dashed lines).

In our prior work, we developed a Conversion Likelihood Estimator (CLE) \cite{cle} model for optimal signal assignment across two primary placements, the Urgency placement (displaying up to 1 signal) and the Conversational placement (displaying up to 2 signals simultaneously). The CLE model outputs an assignment tuple of 3 signals total across both placements and has generated significant GMB uplift in the latest A/B test. However, this approach has a critical limitation, it does not incorporate user preferences or detailed item characteristics, resulting in deterministic signal assignments whenever a specific set of signals qualify.

To address this limitation, we introduce Stageboost, a novel two-stage XGBoost-based approach that enables personalized signal assignment in addition to  explicitly modeling conversion uplift. In the first stage (M1), the model learns to predict the baseline probability of conversion for a given item-user pair using contextual features such as item attributes (price, category, seller reputation) and user characteristics (platform, vertical, buyer segment). In the second stage (M2), the model learns to estimate the incremental uplift in conversion probability attributable to showing each specific signal, effectively learning which signal provides maximum marginal value on top of the baseline prediction.

\subsection{Related Work}

In our prior work, we developed the Conversion Likelihood Estimator (CLE) \cite{cle} as the first machine learning approach for VI signals assignment. The CLE approach directly calculates signal conversion rates within each qualification set and ranks signals based on these empirical rates, provided their differences (representing uplift) are statistically significant. For example: Based on controlled random signal assignments, if we see that when "Purchased n times" and "In n carts" qualifies and we have a higher conversion for "Purchased n times", then this means for qualification set of <"Purchased n times" and "In n carts">, CLE will always recommend "Purchased n times". 

While the CLE achieved impressive results (0.28\% increase in GMB), it has fundamental limitations. First, it provides no personalization: identical qualification sets always receive identical signal assignments regardless of item characteristics, user preferences, or contextual factors. Second, it suffers from data sparsity for rare qualification tuples, as each tuple requires sufficient historical observations to establish statistically significant conversion rate differences. These limitations motivated the development of Stageboost, which explicitly models the relationship between item/user features and signal effectiveness.

There are several other ML approaches that have been applied to similar problems. Booking.com’s paper \cite{Goldenberg2020FreeLunch} presents uplift modeling methods used in the context of coupon offering optimization. Reinforcement Learning (RL) has been studied for reducing cart abandonment \cite{rlcart}, where agents choose actions (cart regularization, discount offers , scarcity messages) based on feature vectors. While contextual bandits \cite{sawant2018contextual}  present promising opportunities for online learning and adaptation, we initially focused on supervised learning approaches to facilitate faster go-to-market, with plans to explore RL \cite{Li2018RLUplift} in future work.

Causal estimation \cite{pearl2016causal, Hernan2020WhatIf} and uplift modeling \cite{Jeong2020CausalML, Gutierrez2020CausalInference} approaches have been successfully applied to similar personalization problems. However, these methods proved less effective in our setting due to the extremely small uplifts (order $10^{-4}$) and high correlation among signals. This resulted in developing Stageboost, which combines the simplicity of gradient boosting with explicit two-stage modeling of base conversion and signal-specific uplift.

Our decision to collect training data via randomized controlled trials (RCTs) rather than purely observational methods was informed by \cite{Gordon2019AdvertisingMeasurement}, which compares observational studies with RCTs for measuring advertising effects.

Stageboost relates to the uplift modeling literature, which focuses on estimating heterogeneous treatment effects, predicting how different interventions affect outcomes for different units. Traditional uplift approaches include two-model methods that train separate models for treatment and control groups, meta-learners that learn treatment effect heterogeneity directly, and tree-based methods like causal forests. Stageboost adopts a two-stage architecture conceptually similar to these approaches, M1 predicts baseline conversion (analogous to control outcome), while M2 estimates signal-specific uplift. However, our single-model XGBoost \cite{chen2016xgboost} formulation with differential feature weighting is tailored specifically to the computational constraints and data characteristics of eBay's production environment, where strict serving latency and large-scale training data are critical requirements. To our knowledge, Stageboost is the first to apply a unified single model two-stage gradient boosting approach to the VI signals assignment problem at scale, building on and extending ideas from existing uplift modeling and meta-learner approaches.

\section{Problem Statement}
\label{sec:problem}
Let $\mathcal{X}$ denote the feature space of item-user pairs, and let $\mathcal{S} = {s_1, s_2, \ldots, s_N}$ represent the set of all possible signals. For a given VI impression, let $x \in \mathcal{X}$ denote the context vector containing item and user features, and let $Q \subseteq \mathcal{S}$ denote the set of qualified signals for that impression (signals that pass eligibility preprocessing). For placement $i$ with $k_i$ display slots, the goal is to select and rank a subset $R \subset Q$ where $|R| = \min(k_i, |Q|)$ to maximize the expected conversion probability.

Formally, we aim to learn a ranking function $f: \mathcal{X} \times \mathcal{S} \rightarrow \mathbb{R}$ that assigns a score to each (context, signal) pair, such that for qualified signals $s_j \in Q$, the selected signals maximize:

\begin{equation}
 R^* = \arg\max_{R \subset Q, |R|=k_i} \mathbb{E}[Y \mid x, R]
\end{equation}

where $Y \in {0, 1}$ is the binary conversion indicator (e.g., Purchase event).

\section{Methodology}
\subsection{Overview}\label{sec:methodology_overview}

Stageboost is a two-stage gradient boosting approach designed to personalize VI signal assignment by explicitly modeling both base conversion probability and signal-specific uplift. The model consists of two distinct stages trained within a single XGBoost ensemble. M1 (Baseline Conversion Model) learns to predict the baseline probability of conversion for a given item-user pair in the absence of signal-specific information. M2 learns to estimate the incremental uplift in conversion probability attributable to displaying each specific signal. 

At inference time, for a VI impression with context $x$ and qualification set $Q = {s_{j_1}, \ldots, s_{j_m}}$, we rank signals by their predicted conversion probabilities:

\begin{equation}
  \text{rank}(s_j \mid x) = p_2(x, s_j) \quad \forall s_j \in Q
\end{equation}

where $p_2(x, s_j)$ is the probability of conversion from M2.

The top-$k_i$ signals are selected for display:
\begin{equation}
  R^* = \text{top-}k_i\left({(s_j, p_2(x, s_j)) : s_j \in Q}\right)
\end{equation}

This formulation enables personalized signal assignment that adapts to both user context and item characteristics, maximizing the expected conversion probability for each individual impression.

\begin{figure*}[t]
    \centering
    \includegraphics[
        trim={0cm 0.1cm 0cm 0.1cm},
        clip,
        width=0.7\linewidth
    ]{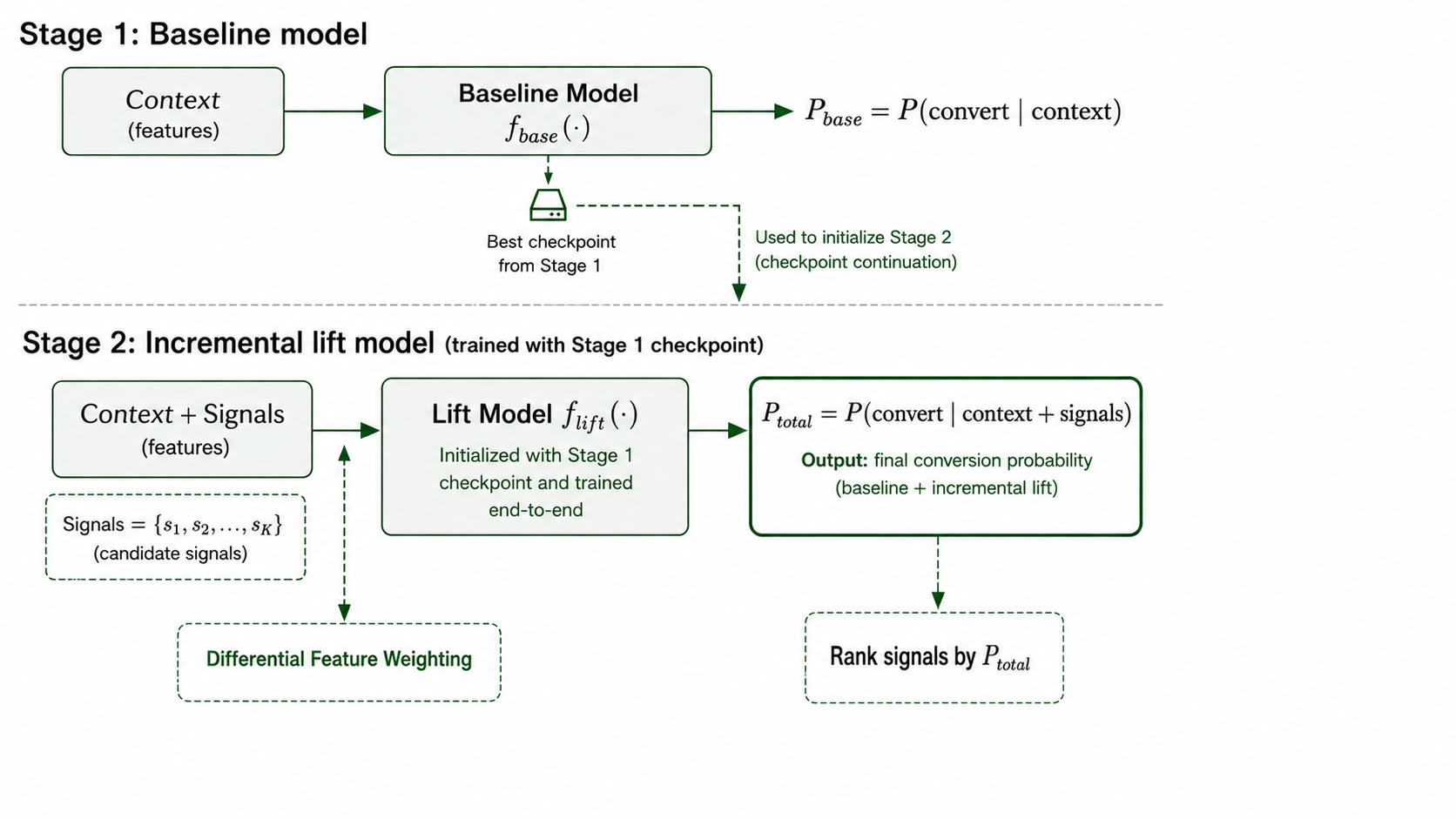}
    \caption{Stageboost Architecture}
    \label{fig:arch}
\end{figure*}

\subsection{Data Preparation}
Training data for Stageboost is collected through Randomized Controlled Trials (RCTs) conducted on eBay's production VI page. Unlike purely observational data collection, which can suffer from selection bias and confounding, RCTs provide unconfounded experimental evidence of signal effectiveness by randomly assigning VI impressions to treatment groups where signals are displayed according to different assignment strategies. Randomization occurs within qualification sets, for example if signals {"In x Carts", "Purchased y times", "Viewed z times"} qualify for a given impression, the treatment group randomly selects which signal(s) to show with uniform probability. This ensures that every qualified signal has a chance of being assigned regardless of item/user features, eliminating selection bias and providing balanced training data across all qualified signals. We are unable to use CLE production data to train stageboost as it would introduce multiple biases that would prevent the model from learning effective personalization. For a given qualification set $Q = {\text{In x Carts},\ \text{Purchased y times},\ \text{Viewed z times}}$, CLE always assigns the same ranking, say signal  $${\text{Purchased y times}}$$ ranked first because it historically had the highest conversion rate for this qualification set. This means we only observe conversions for signal "Purchased y times" when qualification set $Q$ appears and never observe what would have happened if we had shown signal "In x Carts" or "Viewed z times". Randomization ensures uniform exploration, providing balanced training data across all signals regardless of their historical performance. We use 2 weeks of data collected using randomization for training the model and ~1 week of data for offline evaluations.

\subsection{Training}
Stageboost employs a single XGBoost classifier trained with specialized techniques to balance the dual objectives of base conversion prediction (M1) and signal-specific uplift estimation (M2). The training  process leverages differential feature weighting and strategic feature subsampling to ensure the model learns both stages effectively within a unified gradient boosting framework.

Rather than training separate models for M1 and M2, we train a single XGBoost classifier on the full feature set including both item/user features and signal assignment indicators. The key innovation is the use of feature importance weights to control which features dominate learning at different stages of boosting. Specifically, we construct a feature weight vector where signal assignment indicator features (e.g., binary indicators for which signal was shown) receive significantly higher importance weights, while all other features (item characteristics, user context, engagement metrics) receive substantially reduced weights. This extreme differential weighting between signal indicators and context features—ensures that early boosting rounds primarily learn base conversion patterns from item/user features (M1 behavior), while later rounds focus almost exclusively on signal-specific uplift patterns (M2 behavior). XGBoost's feature importance weighting mechanism applies these weights when computing split gain, effectively making context features orders of magnitude less attractive for splitting than signal indicators once M1 patterns are established. Additionally, we apply feature subsampling per node controlled by a delta parameter, which sets XGBoost's per-node column sampling parameter to min(1.0, (K + D) / M) where K is the number of signal indicators and M is the total number of features. Smaller D (delta) values result in aggressive feature subsampling and controls the number of non signal features from M1 that will appear in a split.

To ensure M1's base conversion predictions remain stable and high-quality, we employ sequential training controlled by a configuration flag. In the first phase, we train M1 using only item and user features (excluding shown signal indicators). In the second phase, we continue training from M1's checkpoint by initializing M2 with the pretrained M1 model and fitting on the expanded feature set that includes both the original context features and signal assignment indicators. Typical M2 hyperparameters differ from M1: a lower learning rate for more careful refinement, increased maximum depth to capture signal-context interactions, and the feature weighting and subsampling schemes described above.

\subsection{Evaluation Metrics}
Offline evaluation metrics are essential for comparing model variants and tuning hyperparameters prior to costly online experimentation. However, standard classification metrics like AUC, precision, and F1-score prove inadequate for the signal assignment task. Standard metrics can be misleading, a model may achieve good classification accuracy while producing negative uplift by consistently ranking lower-converting signals above higher-converting alternatives. Since signals have nearly identical conversion rates, traditional metrics cannot distinguish between a model that makes good signal assignment decisions versus one that systematically chooses suboptimal signals. 

To address these limitations, we use a specialized offline metric developed in our prior work\cite{cle} that directly estimates the expected conversion rate if a model's predicted signal assignments were deployed in production. The key insight is to leverage the randomized experimental data to compute counterfactual outcomes. For each model prediction, we estimate "what would the conversion rate be if we actually showed this signal?" by examining instances in the randomized data where the model's prediction matches what was randomly assigned. Since randomization ensures these matching instances are representative, we can estimate the performance of the model's full assignment strategy.

For a given model M, we estimate its expected performance by partitioning test impressions based on the model's predictions (e.g., impressions where M recommends signal "In n carts" vs. signal "Purchased n times"). Within each partition, we calculate the conversion rate using only instances where the randomized experiment actually showed that signal. This gives us an unbiased estimate of what would happen if we deployed Model's recommendations.                                                                                         Formally, let $V^M_i$ denote the set of impressions where model M predicts signal $s_i$. The counterfactual conversion rate for this group is estimated using the adjustment formula:

\begin{equation}
\begin{split}
P(Y=1 \mid T=s_i, V^M_i) = {} & \sum_z P(Y=1 \mid T=s_i, Z=z, V^M_i) \\
& \cdot P(Z=z \mid V^M_i)
\end{split}
\end{equation}

where $Y$ is the conversion indicator, $T$ is the signal assignment, and $Z$ represents the qualification set (which signals qualified for display). The overall estimated conversion rate if M were deployed is:

\begin{equation}
  \hat{c} = \sum_i \frac{|V^M_i|}{N} \cdot P(Y=1 \mid T=s_i, V^M_i)
\end{equation}

where $N$ is the total number of test impressions. This weighted average represents the expected conversion rate under M's assignment strategy, accounting for qualification set confounding. Full mathematical details are provided in our previous CLE work \cite{cle}. 

\section{Empirical Results}
We evaluate two variants of the StageBoost framework that differ in the number of predictors used. The Lean StageBoost variant follows the same architectural design as StageBoost but uses a reduced feature set to satisfy the production latency budget. In contrast, the full StageBoost model incorporates a larger feature set and achieves higher estimated uplift, albeit at the cost of increased inference latency. Given the strict serving latency requirements, Lean StageBoost was selected for online deployment.

Table~\ref{tab:uplift} summarizes the counterfactual evaluation results. The existing CLE model serves as the production baseline, and all reported uplifts are measured relative to random signal assignment using the counterfactual estimation framework described in the previous section. While the full StageBoost model achieves the highest estimated purchase uplift, Lean StageBoost provides a favorable trade-off between effectiveness and latency, making it suitable for production deployment.

\begin{table}[htbp]
    \centering
    \small
    \setlength{\tabcolsep}{6pt}
    \caption{Counterfactual uplift measurement for different models. CLE denotes the production baseline.}
    \label{tab:uplift}
    \begin{tabular}{lccc}
    \toprule
    \textbf{Model} & \textbf{Purchase} & \textbf{Latency} \\
    \midrule
    CLE (Baseline) & 0.109\% & Within Budget \\
    StageBoost & 0.255\% & Exceeds Budget \\
    Lean StageBoost & 0.144\% & Within Budget \\
    \bottomrule
    \end{tabular}
\end{table}

To validate the offline gains, we conducted a large scale online A/B experiment using Lean StageBoost, covering both Urgency and Conversational placements simultaneously. Each placement was treated independently and for the lean stageboost variant, the top 3 signals (1 urgency and 2 conversational) were selected based on conversion probability from the qualified candidate set. At the overall marketplace level, the experiment resulted in a **0.08\% increase in Gross Merchandise Bought (GMB)** , although it was not statistically significant (based on p-value). More pronounced improvements were observed in one of the primary focus categories for this work. In this category, the proposed approach was particularly effective at promoting higher-value purchases, resulting in a 0.58\% increase in GMB and 0.41\% increase in Average Selling Price (ASP), indicating that the optimized placement strategy helped users discover and purchase higher-value items.

\begin{table}[htbp]
\centering
\small
\caption{Online A/B experiment results on live traffic. $\star$ denotes statistical significance.}
\label{tab:ab_results}
\begin{tabular}{lcc}
\toprule
\textbf{Metric} & \textbf{Overall} & \textbf{Focus Category} \\
\midrule
GMB & $+0.08\% \pm 0.28\%$ & $+0.58\% \pm 0.50\% \, \star$ \\
ASP & -- & $+0.41\% \pm 0.38\%$ $\star$ \\
\bottomrule
\end{tabular}
\end{table}

Based on these positive online results, the proposed ranking system has been launched in production. To ensure that signals are displayed only when meaningful choices exist, the model is invoked only when the number of eligible signals exceeds the number of available placement positions.

\section{Conclusion and Future Work}
In this paper, we introduced Stageboost, a novel two-stage XGBoost-based approach for personalized signal assignment on eBay's View-Item page. Unlike the previous CLE model which provided deterministic assignments, Stageboost explicitly models both base conversion probability and signal-specific uplift through a unified gradient boosting framework with differential feature weighting. This enables personalized signal recommendations that adapt to item characteristics and user context.

Our approach addresses several critical technical challenges inherent to this problem: extremely small uplift magnitudes (0.0001), severe data sparsity in rare qualification sets, and high feature correlation among signals. By collecting training data through RCTs rather than observational methods, we ensure unbiased training signal coverage across all qualified signals through randomized exploration. The specialized offline evaluation metric based on counterfactual conversion rate estimation proved essential for model selection, where traditional classification metrics failed to distinguish between models with significantly different business impact. Given strict production latency constraints and observations based on online experimentation, we have Lean Stageboost model in production. We acknowledge that personalized urgency and scarcity signals optimized for engagement may influence user autonomy. To address this, we monitor signal validity, buyer-segment-level outcomes, and post-purchase indicators as part of our ongoing production evaluation.

Future work includes exploring reinforcement learning approaches for online adaptation, incorporating additional user features, and extending the framework to additional placements on the VI page.

\bibliographystyle{ACM-Reference-Format}
\bibliography{sigproc}

\end{document}